\documentclass[aip, apl, reprint, amsmath, amssymb]{revtex4-2}

\usepackage{graphicx}
\usepackage[caption=false]{subfig}
\usepackage{comment}

\usepackage{url}
\usepackage{dcolumn}
\usepackage{bm}
\usepackage[utf8]{inputenc}
\usepackage[T1]{fontenc}
\usepackage{mathptmx}
\usepackage{etoolbox}
\usepackage[english]{babel}
\usepackage{physics}

\usepackage{pdfpages}

\makeatletter
\AtBeginDocument{\let\LS@rot\@undefined}
\makeatother

\newcommand{\pt}[1]{\left(#1\right)}

\makeatletter
\def\@email#1#2{%
  \endgroup
  \patchcmd{\titleblock@produce}
   {\frontmatter@RRAPformat}
   {\frontmatter@RRAPformat{\produce@RRAP{#1\href{mailto:#2}{#2}}}\frontmatter@RRAPformat}
   {}{}
}%
\makeatother

\begin{document}

\title[]{Phase-correlation-free quantum key distribution source operating at gigahertz rates}

\newcommand{\Geneva}{Department of Applied Physics, University of Geneva, CH-1205 Geneva, Switzerland}
\newcommand{\VigoA}{Vigo Quantum Communication Center, University of Vigo, Vigo E-36310, Spain}
\newcommand{\VigoB}{Escuela de Ingeniería de Telecomunicación, Department of Signal Theory and Communications, University of Vigo, Vigo E-36310, Spain}
\newcommand{\VigoC}{AtlanTTic Research Center, University of Vigo, E-36310, Spain}
\newcommand{\IDQ}{ID Quantique SA, CH-1227 Geneva, Switzerland}

\author{S. Kumar}
\thanks{These authors contributed equally to this work.}
\email{amarcomini@vqcc.uvigo.es}
\affiliation{\Geneva}

\author{A. Marcomini}
\thanks{These authors contributed equally to this work.}
\email{shashank.kumar@unige.ch}
\affiliation{\VigoA}
\affiliation{\VigoB}
\affiliation{\VigoC}

\author{L. Millet}
\affiliation{\Geneva}
\affiliation{\IDQ}

\author{T. Taher}
\author{A. Cavalié}
\author{R. Houlmann}
\author{D. Cabrerizo}
\affiliation{\Geneva}

\author{G. Boso}
\affiliation{\IDQ}

\author{M. Curty}
\affiliation{\VigoA}
\affiliation{\VigoB}
\affiliation{\VigoC}

\author{R. Thew}
\author{B. Korzh}
\affiliation{\Geneva}

\begin{abstract}
Phase randomization is essential for the security of practical decoy-state quantum key distribution (QKD) systems. Commonly, implementations rely on laser sources which are either actively phase-randomized, or gain-switched. However, at high repetition rates these show correlations, which can ultimately compromise security and performance. We present a 1.25 GHz phase-randomized QKD source based on a super-luminescent light emitting diode (SLED) operating in the C-band as a compact and cost-effective alternative. The source generates $\sim100$ ps optical pulses with $400$ ps pulse-to-pulse separation, compatible with high-speed time-bin encoding. Interferometric measurements demonstrate $>99\%$ visibility between adjacent time bins, confirming strong first-order coherence within the same quantum signals, while the spontaneous-emission-driven nature of the SLED ensures intrinsic global phase randomization between adjacent signals. This work establishes a scalable SLED-based platform for high-speed prepare-and-measure QKD systems.
\end{abstract}

\maketitle

Quantum key distribution (QKD) enables long-term security immune from store-now-decrypt-later attacks, overcoming fundamental limitations of classical cryptographic systems and ultimately guaranteeing information-theoretic secrecy of communications without any computational assumption~\cite{bennett1984, ekert1991b, shor2000, mayers2001, gisin2002, lo2014, pirandola2020b}. To date, successful implementations have been demonstrated in realistic scenarios~\cite{aquina2025}, both at urban and intercity levels~\cite{zheng2026}, over classical fiber networks~\cite{pittaluga2025,wu2025}, as well as over satellite links~\cite{liao2017b,yin2020b,chen2021b}. Nevertheless, its widespread application is still conditioned on the resolution of practical limitations, namely its poor performance at medium-to-long distances and the complexity of guaranteeing the implementation security of real, imperfect devices~\cite{scarani2009b, xu2020, brazaola-vicario2024, zapatero2025b}. 

The former limitation can be addressed by employing quantum repeaters~\cite{briegel1998b, azuma2023b, tittel2025}, or increasing the throughput of standard prepare-and-measure (P\&M) protocols~\cite{grunenfelder2023,li2023}. Another breakthrough in this context has been the introduction of twin-field QKD~\cite{lucamarini2018b} which, relying on single-photon interference, manages to effectively double the achievable distance with respect to other protocols in similar conditions.
As for implementation security, one needs to accurately understand and characterize the behavior of practical systems, operating them in conditions which minimize security threats. In fact, when the assumptions of security proofs are violated, so are their claims, ultimately compromising the security of the key~\cite{zapatero2025b, tupkary2025, bsi2023}. Such violations can arise either due to active hacking of an adversary~\cite{makarov2024} or due to unnoticed information leakage in additional modes and multi-photon emissions. 
The latter are particularly problematic because of the so-called photon-number-splitting attack~\cite{brassard2000}, which is commonly mitigated by means of the decoy-state technique~\cite{hwang2003,wang2005b,lo2005b,ma2005b,lim2014d,rusca2018b}. Crucially, this approach requires global phase-randomization of the transmitted signals. 

In recent years, notable efforts have been carried out to address security concerns induced by correlations~\cite{pereira2020,pereira2024,agulleiro2025, curras-lorenzo2026}, imperfect phase randomization~\cite{curras-lorenzo2024,marcomini2025c} and general information leakage~\cite{sixto2025} for protocols relying on weak coherent light pulses (WCPs), typically employed for their ease of calibration and relatively low cost. Although these theoretical efforts practically close security loopholes, the penalty in performance is considerable. Moreover, they might work for only a restricted class of attacks ~\cite{curras-lorenzo2024}, whereas the goal should always be to suppress imperfections to the greatest extent possible.

To solve the phase randomization problem in P\&M QKD protocols, we propose the use of a superluminescent light-emitting diode (SLED), as an alternative to standard lasers. Since the light emitted by SLEDs is generated via amplified spontaneous emission, it displays inherent ultra-short coherence, thus phase correlations are natively absent even at time-scales compatible with gigahertz repetition rates. For this reason, SLEDs have already been exploited in the context of quantum random number generation~\cite{chen2017c,yang2021,li2021}, and have been recently proposed as high-speed entropy sources to enhance the phase randomization of laser-based QKD sources~\cite{lo2026}. 

In this Letter, we show how SLEDs can be directly used as a QKD source to prepare high quality states for the time-bin three-state protocol~\cite{boaron2018d}, displaying no phase correlations among successive signals. Subsequently, we generate coherent time-bin signals through the use of a delay-line interferometer, leading to visibilities similar to those achieved with a gain-switched laser. Finally, we investigate the photon statistics of the SLED-based source and provide a secret key rate (SKR) comparison between the transmitter in question and one exhibiting the photon-number statistics of perfect weak-coherent pulses. This work enables a simple and low cost solution for QKD transmitters, while fulfilling the assumptions of the underlying security proofs. 

\begin{figure}[b!]
\includegraphics[width=.99\columnwidth]{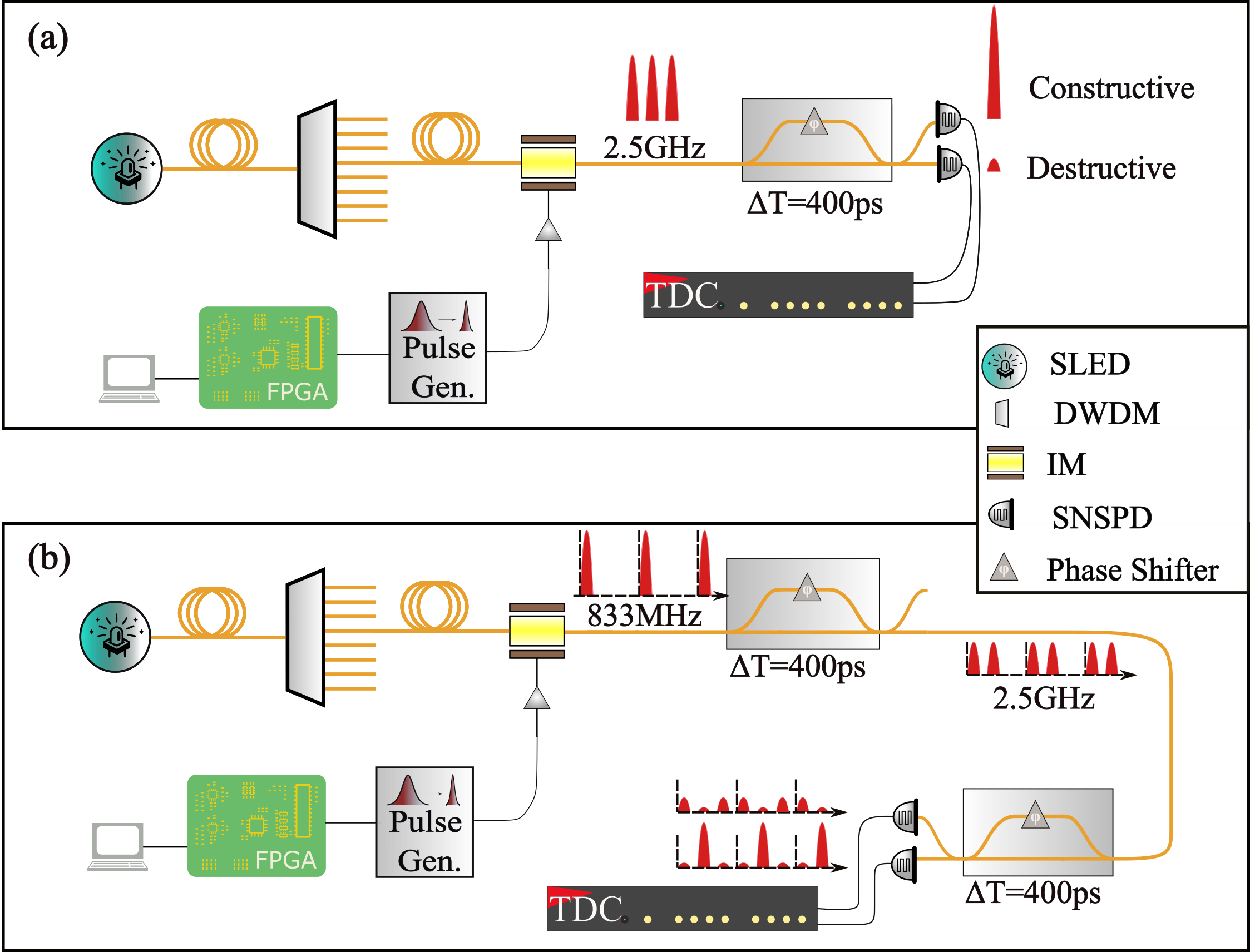}
\caption{Experimental setup for phase-correlation measurements.
(a) Pulse-carved signals at 2.5 GHz are sent to a matched delay-line interferometer (DLI) to test the phase coherence between neighboring pulses by monitoring for constructive and destructive interference at the outputs. The relative phase is controlled by a voltage bias.
(b) Pulses carved at 833 MHz are sent through a first DLI to generate time-bin signals (i.e., early and late time bins with 400 ps separation). A second matched DLI recombines the two bins to measure their relative phase coherence. The outputs are detected using superconducting nanowire single-photon detectors (SNSPDs) connected to a time-to-digital converter (TDC).}
\label{fig:block1}
\end{figure}

The transmitter is based on a C-band SLED ( Exalos EXS210071) with an emission bandwidth of approximately 60 nm full-width at half-maximum (FWHM) --- see Supplementary Material (SM) for spectral data. The SLED output is spectrally filtered using a dense wavelength division multiplexer (DWDM) with an approximate bandwidth of 0.68 nm and externally pulse-carved using a 20 GHz intensity modulator (IM, Agiltron), driven by FPGA-controlled electrical pulses compressed to approximately 53 ps using a pulse generator (Alnair EPG-210). We first test whether successive carved pulses exhibit phase correlations, which would violate the phase-randomization requirement of decoy-state QKD. For this measurement, the IM is driven at 2.5 GHz and the resulting pulse train is sent through a delay-line interferometer (DLI, Exail MINT) with a matching free spectral range of 2.5 GHz, so that neighboring pulses interfere, as shown in Fig.~\ref{fig:block1}(a).

\begin{figure}[t!]
\includegraphics[width=.99\columnwidth]{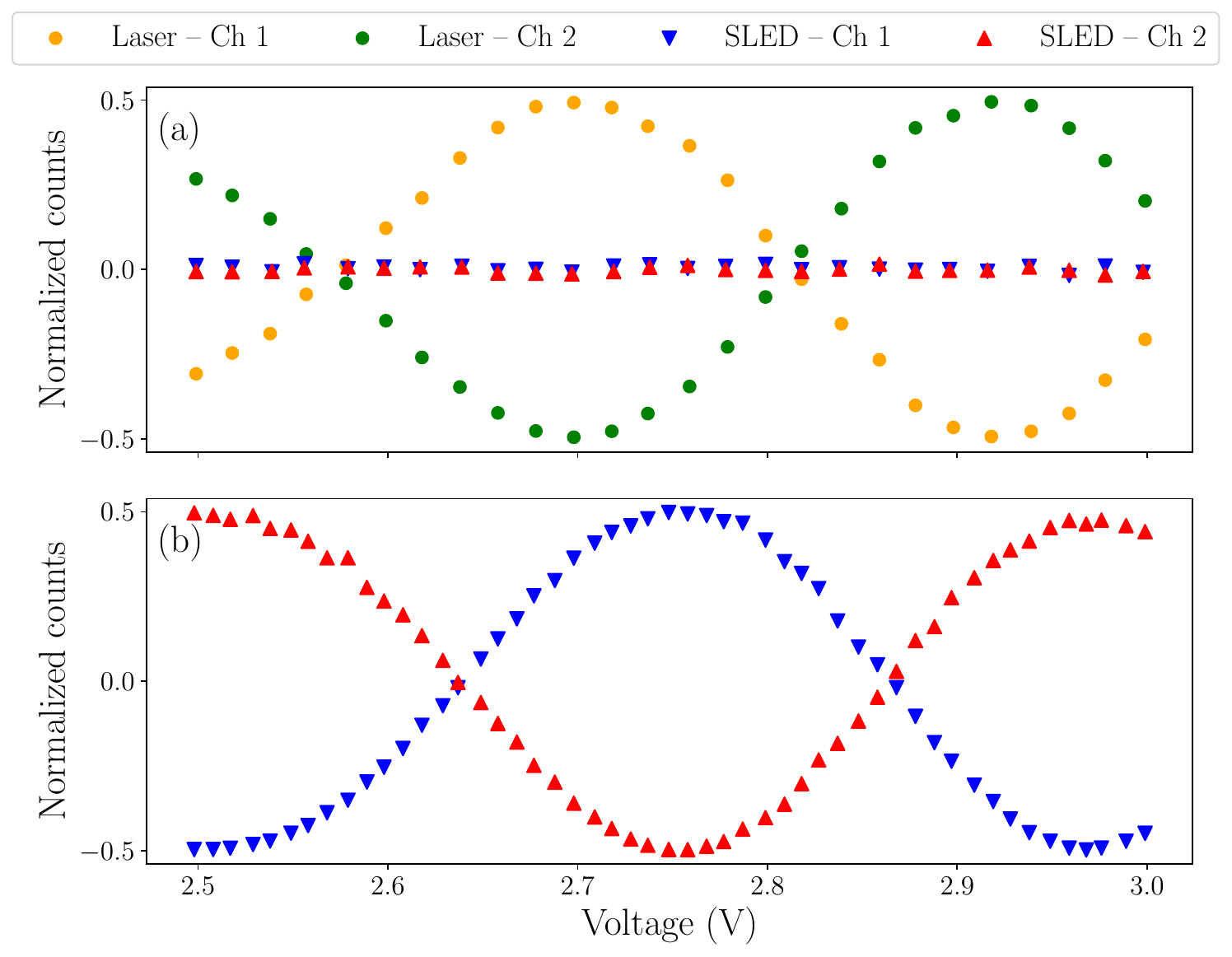}
\caption{Visibility comparison for laser and SLED sources. (a) Nearest-neighbour interference (one-DLI configuration), determining the global phase relation of pulses in different rounds of the protocol. (b) X-basis measurement (two-DLI configuration), corresponding to a measurement of the coherence among time-bins within the same signal.  Counts $N$ are normalized as $N/N_\text{max} - (N_\text{max}+N_\text{min})/2$, being $N_\text{max}$ ($N_\text{min}$) the maximum (minimum) counts of each curve.}
\label{fig:visibility_combined}
\end{figure}

We sweep the relative phase of the DLI by applying a voltage to the thermal phase shifter in one of the interferometer arms. As a reference, the same measurement is also performed using a continuous-wave laser passing through the same optical path. For a phase-correlated source, the detector counts oscillate sinusoidally between the two output ports as the phase is swept. On the contrary, a phase-randomized source shows no phase-dependent interference pattern, with the counts remaining balanced between the two detectors. This is indeed what we observe in Fig.~\ref{fig:visibility_combined}(a), where there is no observable first-order coherence between neighboring pulses for the SLED. Quantitatively, the observed visibilities for the two sources in this setting are computed as $(N_\text{max} - N_\text{min})/(N_\text{max} + N_\text{min})$ assuming Poissonian counting noise. We obtain
\begin{eqnarray}
    V_{\text{Laser}}^{\text{1DLI}} = (98.0\pm0.2)\% \,, \quad V_{\text{SLED}}^{\text{1DLI}} = (1.8\pm0.6)\%\,.
\end{eqnarray}

Next, we demonstrate the ability to create a coherent time-bin signal using the same source. In this case, the pulse-carving rate is reduced to either $833\,$MHz or $1.25\,$GHz and the first DLI converts a parent pulse into a pair of pulses separated by $400\,$ps, as shown in Fig.~\ref{fig:block1}(b). The second DLI, with a matched path delay and a phase control in one arm, overlaps the early and late pulses, allowing their relative phase coherence to be measured. By sweeping the phase modulator voltage in one DLI we observe clear interference, as shown in Fig.~\ref{fig:visibility_combined}(b), which demonstrates that although the global phase is randomized, the relative phase between the early and late bins within each signal remains well defined. When comparing the observed visibilities in this two-DLI configuration, we find
\begin{eqnarray}
    V_{\text{Laser}}^{\text{2DLI}} = (99.4\pm0.1)\% \,, \quad V_{\text{SLED}}^{\text{2DLI}} = (99.1\pm0.2)\%\,.
\end{eqnarray}
This confirms that the SLED source can provide the intra-signal coherence required for preparing and measuring the control-basis state in time-bin QKD.

\begin{figure}
    \centering
    \includegraphics[width=.9\linewidth]{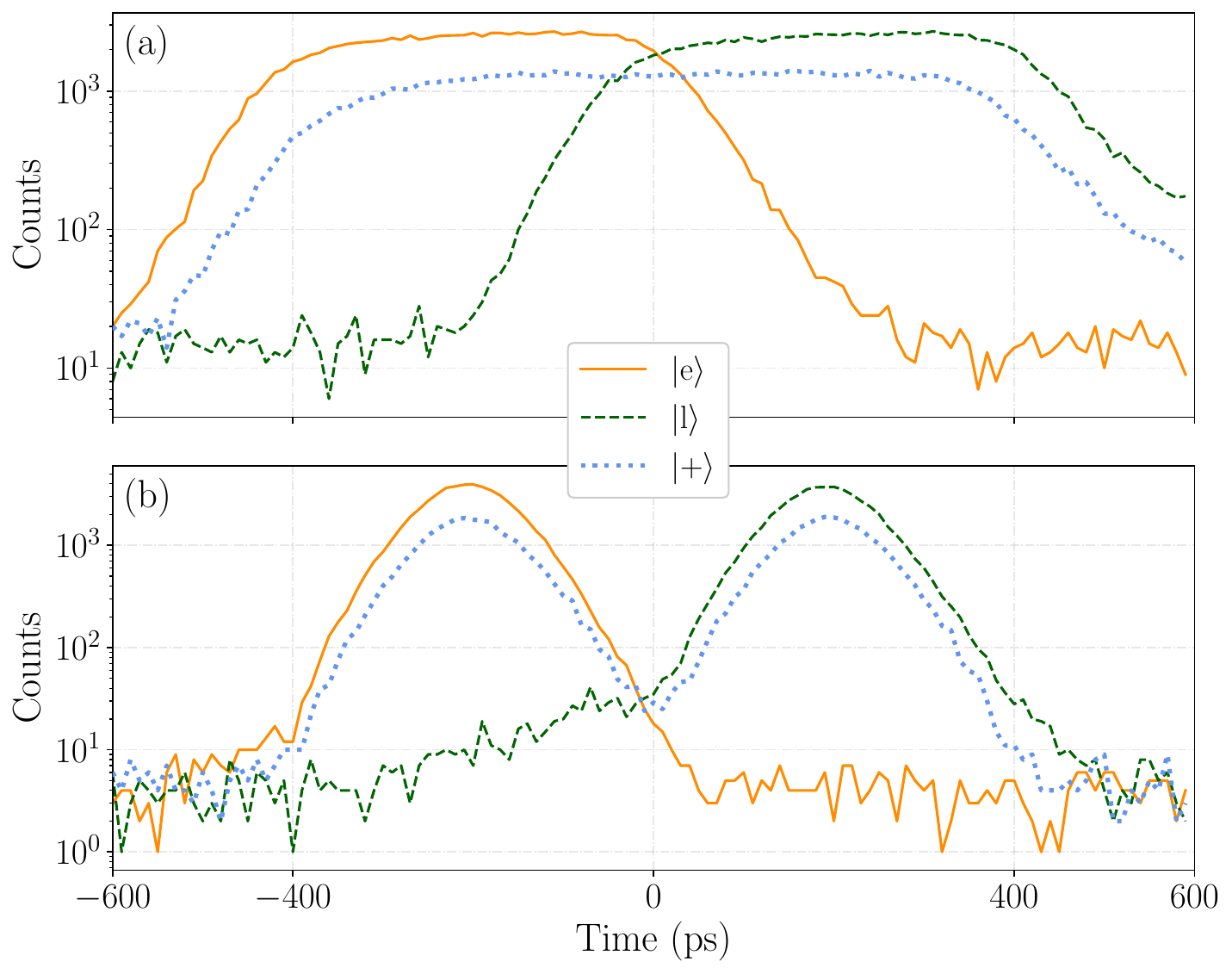}
    \caption{Time-domain measurement of the prepared states. (a) Profile of the state-encoding IM applied to a continuous-wave signal from the SLED. For Z-basis measurements, one time-bin is fully suppressed (with an extinction ratio exceeding 25~dB), while for the X-basis a balanced superposition of the time-bins is encoded. (b) Final states obtained by combining the encoder of panel (a) with pulse-carving. The resulting pulses have a full-width half-maximum of about 120 ps.}
    \label{fig:timedomain}
\end{figure}

We further investigate whether early ($\ket{e}$), late ($\ket{l}$), and superposition ($\ket{+}$) states could be prepared with a sufficiently high extinction ratio for decoy-state time-bin QKD. To do this, the pulse carving rate is set to $1.25\,$GHz, that is, the intended signal generation rate. After passing the first DLI, a second IM is used for state preparation by selecting either the early, late, or superposition state (see SM for the experimental setup). This second modulator is driven by a high-speed digital-to-analog converter (DAC) of the FPGA (AMD ZCU216), providing multi-level signals with a pulse width of 400~ps, as shown in Fig.~\ref{fig:timedomain}(a). Figure~\ref{fig:timedomain}(b) shows the corresponding optical time-bin states when the two modulators are used together. For the early and late states, an extinction ratio of approximately 25 dB is achieved between the occupied and suppressed time bins, confirming that the SLED-based transmitter can achieve a low quantum bit error rate (QBER) in the key-generation basis. As expected, the superposition state exhibits approximately 3 dB lower intensity in each time bin, consistent with an equal distribution of the total intensity between the early and late time bins.

\begin{figure*}[t]
    \centering
    \subfloat{
        \includegraphics[width=0.49\textwidth]{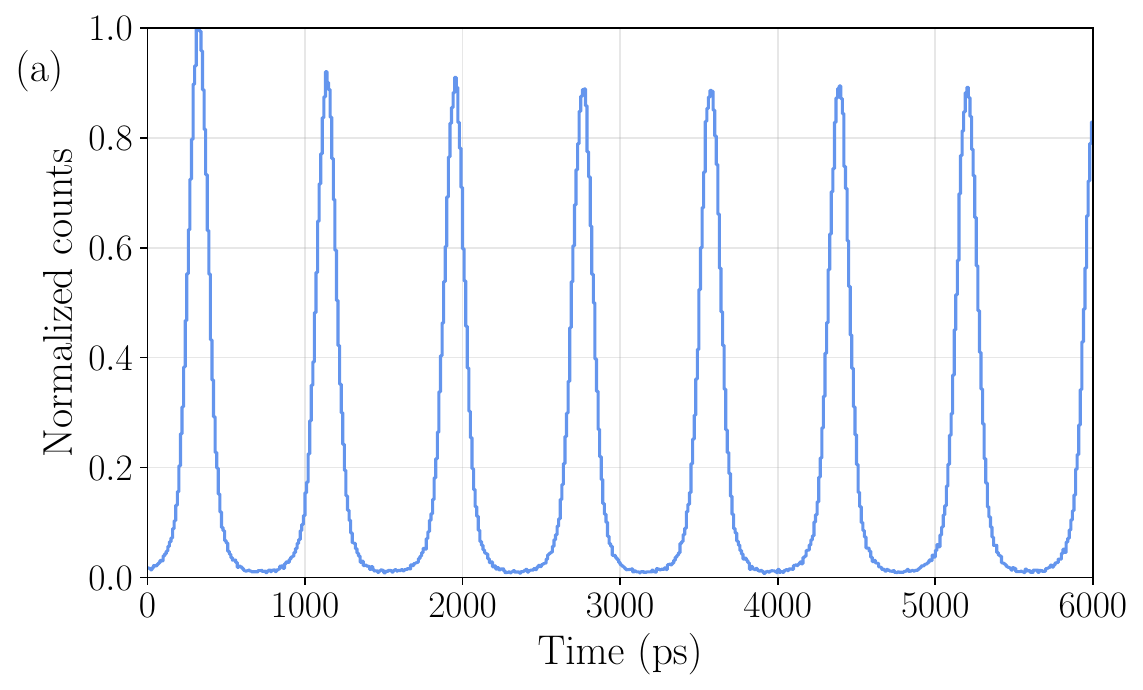}}
    \hfill
    \subfloat{
    \includegraphics[width=0.49\textwidth]{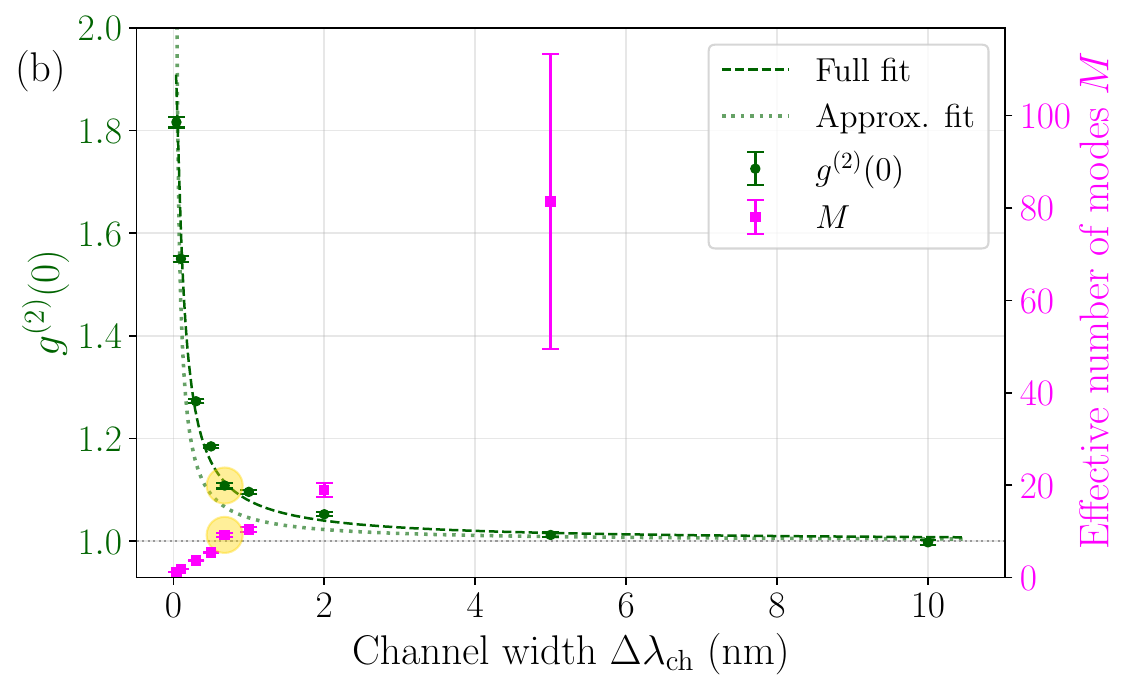}}
\caption{(a) Photon counts versus time with the HBT experiment for the SLED source filtered through the DWDM channel. The primary peak corresponds to zero-time coincidences, and is purposely delayed to capture the full peak shape. The $N_{\mathrm{side}} = 6$ secondary peaks occur periodically, in accordance with the source pulsing rate. The $g^2(0)$ value is obtained through Eq.~\ref{eq:estimate g2} considering an integration window of 100 ps. (b) Scaling of $g^2(0)$ estimates upon changing the channel bandwidth through the OSA tunable filter, with the corresponding fit given by Eq.~\ref{eq: g2 vs M} with $M=\gamma$ (approximate fit) and $M=M(\gamma)$ (Eq.~\ref{eq:M(gamma)}, full fit), where $\gamma \propto \Delta\lambda_\mathrm{ch}$. For each measurement, the corresponding effective mode number $M$ is derived through Eq.~\ref{eq: g2 vs M}. The highlighted data points correspond to the $g^2(0)$ measurement for the nominal 0.68 nm wide DWDM in panel (a) and its corresponding number of modes. }
\label{fig: g2 measurements}
\end{figure*}


Since the security of QKD relies on the emission of single-photon states, it requires a careful characterization of the source output. In fact, while the decoy-state method can be applied to QKD setups with arbitrary photon-number distributions~\cite{foletto2022}, it is important to relate the emission probability of $n-$photon pulses to experimental quantities that can be conveniently accessed in practice. 
Notably, the photon-number emission statistics of an SLED producing amplified spontaneous emission have been widely studied~\cite{wong1998,pietralunga2003,yang2021}. The probability of detecting $n$ photons by a photo-detector over an average detection time $T_{\mathrm{det}}$, given a mean photon number $\bar{n}$, when the source emits $M$ independent optical modes, has been shown to follow the $M$-fold degenerate Bose-Einstein distribution
\begin{eqnarray}\label{eq:BoseEinsteinDistribution}
    P_{BE} \pt{n;\bar{n},M} := \frac{\Gamma\pt{n+M}}{\Gamma\pt{n+1}\Gamma\pt{M}}\pt{1+\frac{M}{\bar{n}}}^{-n}\pt{1+\frac{\bar{n}}{M}}^{-M}
\end{eqnarray}
where $\Gamma\pt{x}$ denotes the Gamma function. The effective number of temporal modes $M$ can be deduced by experimentally measuring the degree of second order coherence, for which it holds~\cite{loudon2000,mandel1996}
\begin{eqnarray}\label{eq: g2 vs M}
    g^{\pt{2}}\pt{0} = 1+\frac{1}{M}.
\end{eqnarray}

For the particular case of chaotic light with a Gaussian power spectral density, assuming a balanced amount of photons per mode, the above expression yields an explicit formula in terms of the optical channel bandwidth $B_{\text{opt}}$ and the electrical bandwidth of the receiver $B_{\text{el}}:=1/T_{\mathrm{det}}$. By letting $\gamma:=B_{\text{opt}}/B_{\text{el}}$, in this case we have~\cite{pietralunga2003}
\begin{equation}\label{eq:M(gamma)}
M(\gamma) \;=\;
    \frac{s\,\pi\gamma^{2}}
         {\pi\gamma\,\mathrm{erf}\!\left(\sqrt{\pi}\,\gamma\right)
          -\left[1-e^{-\pi\gamma^{2}}\right]},
\end{equation}
where $s$ denotes the polarization degeneracy ($s=1$ for polarized amplified spontaneous emission, $s=2$ for unpolarized) and $\text{erf}(x)$ denotes the error function. Note that $M(\gamma) \simeq \gamma$ for $\gamma\gg 1$.

For practical QKD applications, the broad spectrum of the SLED needs adequate filtering to match the requirements of standard fiber and detectors. In the following, let $\Delta\nu_\mathrm{ch} = B_{\mathrm{opt}}$ denote the filtered channel bandwidth in Hz, and $\Delta\lambda_\mathrm{ch}$ the corresponding value in nm. If Alice sends optical pulses of width $T_\mathrm{pulse} = T_\mathrm{det}$, then the effective number of temporal modes emitted is given by Eq.~\ref{eq:M(gamma)}, with $\gamma := \Delta\nu_\mathrm{ch}\,T_\mathrm{pulse}$~\cite{mandel1996}. Note that both parameters are tunable when designing the protocol.

We infer $M$ by measuring $g^{\pt{2}}(0)$ for different filter widths using a Hanbury Brown-Twiss (HBT) interferometer in the pulsed regime~\cite{mandel1996} (experimental setup in SM).
The coincidence histogram acquired on one output port of the beam splitter displays a series
of peaks separated by the laser repetition period $T_\mathrm{source} = 800$~ps (corresponding to the $1.25\,$GHz modulation, see Fig.~\ref{fig: g2 measurements}(a)), and $g^{(2)}(0)$ is estimated as the ratio of the area of the principal peak ($A_0$, corresponding to zero-delay coincidences) to the mean area of the $N_\mathrm{side}$ lateral peaks $\bar{A}_\mathrm{side}$~\cite{dynes2018}:
\begin{equation}\label{eq:estimate g2}
    g^{(2)}(0) \;=\; \frac{A_0}{\bar{A}_\mathrm{side}},
    \quad
    \sigma_{g^{(2)}} \;=\; g^{(2)}(0)
    \sqrt{\frac{1}{A_0} + \frac{1}{N_\mathrm{side}\,\bar{A}_\mathrm{side}}},
\end{equation}
where $\sigma_{g^{(2)}}$ is the statistical
uncertainty propagated under the assumption $\sigma_A = \sqrt{A}$ (Poissonian shot noise).
Each peak area is integrated over a 100~ps window, 
positioned to maximize the enclosed counts. 
For the 0.68 nm DWDM channel, the analysis yields an estimate of
\begin{eqnarray}\label{eqn: g2 exp}
g^{(2)}(0) = 1.109 \pm 0.006\,.
\end{eqnarray}

To study the dependence of $g^{(2)}(0)$ with 
the spectral filter bandwidth $\Delta\lambda_\mathrm{ch}$, we vary the internal tunable filter of an optical spectrum analyzer (OSA) while keeping the source repetition rate and intensity settings fixed. We fit the observed $g^{(2)}(0)$ data with Eq.~\ref{eq: g2 vs M}, both in the approximate case of $M\approx \gamma \propto \Delta\lambda_\mathrm{ch}$, as well as with the full expression in Eq.~\ref{eq:M(gamma)}. Fitting results are shown in Fig.~\ref{fig: g2 measurements}(b) showing best agreement with the full expression. As $\Delta\lambda_\mathrm{ch}$ is increased, more temporal modes are admitted and $g^{(2)}(0)$ decreases monotonically from $2$ (single-mode thermal bunching) towards $1$ (Poisson-like). 
Contextually, the number of estimated temporal modes grows with $\Delta\lambda_\mathrm{ch}$, in a close-to-linear trend. Despite the small uncertainty on the $g^{(2)}(0)$ estimates, the statistical errors associated to $M$ (derived from Eqs.~\ref{eq: g2 vs M}-\ref{eq:estimate g2}) increase drastically at large bandwidths, as they are inversely proportional to $(g^{(2)}(0) - 1)$. For example, for $\Delta\lambda_\mathrm{ch} = 10$~nm the errors are too large to provide a meaningful estimate (and the corresponding data point is omitted), although uncertainty remains reasonable in the intended operational regime of $\Delta\lambda_\mathrm{ch}<1$~nm.

To assess the feasibility of using an SLED-based transmitter for QKD, we compare the achievable SKR versus a laser-based transmitter for the decoy-state BB84 protocol in the asymptotic scenario~\cite{lo2005b, ma2005b} (note that in this case the results coincide with those of the three-state protocol)~\cite{tamaki2014b}. Detection statistics are simulated using both a Poissonian and multimode-thermal source, while we model a standard noiseless channel of length $L$ as a beam splitter with transmittance
\(\eta = \eta_\text{det} \cdot 10^{-\alpha L/10}\), where \(\alpha =
0.20\;\text{dB/km}\) is the fiber loss coefficient and \(\eta_\text{det} =
0.80\) is the detector efficiency. We consider an active BB84 receiver with dark-count probability \(p_{\mathrm{d}} = 10^{-8}\) per detector per pulse, and randomly assign double clicks. 
The expected gain and QBER for signals with mean photon number ${\mu_k}$ are given by
\begin{align}
    Q_{\mu_k} = \sum_n P(n|\mu_k)\,Y_n = 1 - (1-p_{\mathrm{d}})^{2}\,\left(1-p_{\mathrm{ch}}({\mu_k},\eta)\right), \label{eq:gain}
    \\
    E_{\mu_k} = \sum_n P(n|\mu_k)\, e_n Y_n = \frac{p_{\mathrm{d}} + e_\text{mis}p_\mathrm{ch}({\mu_k},\eta)}{Q_{\mu_k}},
    \label{eq:qber}
\end{align}
where $P(n|\mu_k)$ denotes the probability of sending a $n-$photon pulse when setting a mean photon number $\mu_k$ for a given source model, $Y_n$ and $e_n$ denote respectively the $n-$photon yield and error,
\(e_\text{mis} = 0.5\%\) is the optical misalignment, and
\(p_\mathrm{ch}({\mu_k},\eta)\) denotes the
probability that a pulse delivers at least one photon to the detector. For the Poisson source \(p_\mathrm{ch}({\mu_k},\eta) = 1 - e^{-{\mu_k}\eta}\), while for a
multimode-thermal source with \(M\) modes we have
\(p_\mathrm{ch}({\mu_k},\eta) = 1- (1+{\mu_k}\eta/M)^{-M}\) (see Eq.~\ref{eq:BoseEinsteinDistribution}, with $\bar{n}={\mu_k}\eta$), which reduces to the single-mode
geometric distribution for \(M=1\) and approaches the Poisson limit as
\(M\to\infty\).

The decoy-state analysis employs a three intensity setting (signal~\(\mu_s\),
weak~\(\mu_w\), and vacuum~\(\mu_v=0\))~\cite{ma2005b}.
For each source distribution, we constrain the photon-number yields
\(\{Y_n\}_{n=0}^{N_{\mathrm{cut}}}\) (\(N_{\mathrm{cut}}=20\)) through Eq.~\ref{eq:gain} for each intensity $\mu_k$ (similarly for $\{e_n Y_n\}_n$ through Eq.~\ref{eq:qber}). We can bound the single-photon statistics by solving two linear programs (LPs): one minimizing \(Y_1\)
to obtain a lower bound \(Y_1^L\), and one maximising \(e_1 Y_1\)
to obtain an upper bound \(e_1^U:=\pt{e_1 Y_1}^{U}/Y_1^L\).
The asymptotic SKR per pulse is then given by~\cite{lo2005b}
\begin{equation}
    R \;\geq\; P(0|\mu_s)\,Y_0
             + P(1|\mu_s)\,Y_1^L\,\bigl[1 - h_2(e_1^U)\bigr]
             - f_\text{EC}\,Q_{\mu_s}\,h_2(E_{\mu_s}),
    \label{eq:skr}
\end{equation}
where \(h_2\pt{x}\) is the binary entropy function and \(f_\text{EC} = 1.16\)
is the error-correction efficiency.

\begin{figure}[t]
    \includegraphics[width=.99\columnwidth]{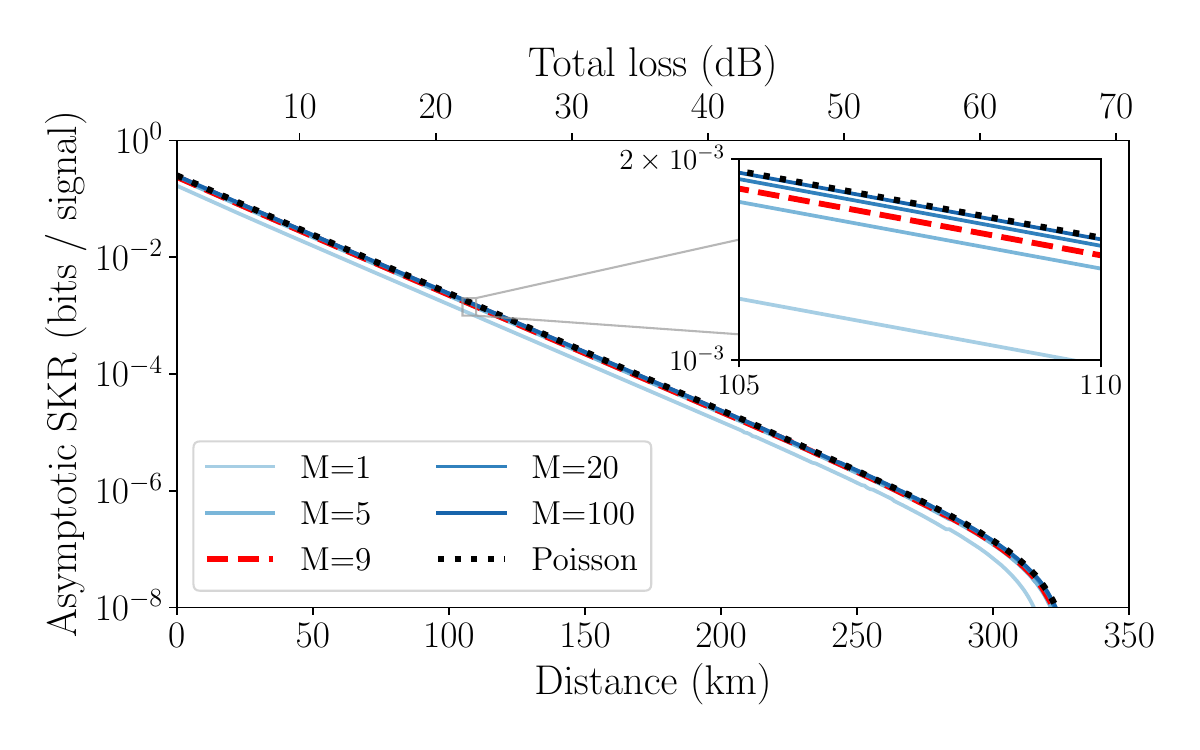}
    \caption{Asymptotic secret key rate obtainable from a phase-randomized laser source (Poissonian statistics, dotted black line), compared with that of multimodal thermal light characteristic of the SLED for different values of the effective mode number $M$ (solid blue lines). The red dashed line corresponds to the conservative estimate of $M$ from the experimental data in Fig.~\ref{fig: g2 measurements}(b). As $M\gg1$, the multimode-thermal distribution approaches the Poissonian, and results correctly converge for $M=100$.}
    \label{fig:SKR}
\end{figure}

At each distance, the intensities $\mu_s$ and $\mu_w$ are optimized to maximize the lower bound on the SKR shown in Fig.~\ref{fig:SKR}. Despite the slightly higher multi-photon emission, results for the SLED source (solid blue lines) are very close in performance to those of a Poissonian source, with a penalty smaller than a factor of two. 
As the number of modes $M$ increases, the performance gradually reaches that of a phase-randomized laser (dotted black line), and the convergence is almost total for $M=100$. The difference is also minimal for a value of $M=9$ (dashed red line), corresponding to the integer part of the number of modes estimated from the experimental result in Eq.~\ref{eqn: g2 exp}. We display this result for comparison purposes, but underline that in an actual implementation the value of $M$ should be properly lower-bounded within the prescription of the applied security analysis. This bound can be conservative without greatly affecting the SKR, as Fig.~\ref{fig:SKR} suggests that the dependence of the SKR on $M$ is minimal for small changes.

We note that results in Fig.~\ref{fig:SKR} explicitly assume that the SLED photon emission statistics follow the distribution in Eq.~\ref{eq:BoseEinsteinDistribution}. One could in principle remove this assumption and simply rely on measurements of autocorrelation functions of different orders, although this will reduce the number of linear constraints, significantly worsening the LP performance~\cite{dynes2018}. Although the photon-number statistics of SLEDs have been extensively studied, an interesting outlook would be the direct characterization of their emission via photon-number-resolving detectors~\cite{stasi2024}. In any case, given the distribution model in Eq.~\ref{eq:BoseEinsteinDistribution}, the practical challenge in a concrete implementation is limited to the estimate of the effective number of modes $M$ (since the mean photon number $\bar{n}$ is actively tuned in the protocol).

We also remark that the curves in Fig.~\ref{fig:SKR} do not assume a specific source repetition rate. Nevertheless, it is well known that for high rates the performance can suffer from correlations. The active IM that is adopted in our setup is vulnerable to intensity correlations in the same way as laser-based schemes, for which dedicated characterization and security proofs apply~\cite{zapatero2021,trefilov2025, navarrete2026}. From this perspective, SLED-based sources do not exhibit any operationally significant difference compared with laser-based counterparts. However, the inherent ultra-short coherence time of the SLED lifts all concerns regarding phase correlations, which can instead induce severe penalties to the performance of laser-based schemes at the GHz level~\cite{kobayashi2014,grunenfelder2020b,marcomini2025c}. Moreover, it provides a key ingredient towards the realization of fully-passive QKD transmitters for decoy-state BB84~\cite{wang2023b,zapatero2023c}, which are inherently robust against hacking attacks targeting optical modulators, as it would remove the primary implementation concern, namely global phase randomization. Note that for these latter schemes intensity correlations are irrelevant as the IM profile is not actively changed at each round.
An additional advantage of SLED-based systems is that they can naturally support DWDM multiplexing. In principle, each wavelength channel could act as an independent QKD channel with its own bit and basis encoding, increasing the total key rate without needing multiple light sources. The main point to confirm in future work is that there are no phase correlations between pulses in different DWDM channels, since such correlations would break the phase-randomization requirement. 


In summary, we investigated an SLED-based source, demonstrating its applicability in GHz-rate time-bin QKD. 
Because SLED emission originates from amplified spontaneous emissions, this source is inherently phase-randomized. This is certified by the fact that interferometric measurements show no appreciable first-order phase correlation between neighboring signals. 
At the same time, the source preserves high phase correlation between the early and late time-bins within the same signal, achieving an interference visibility in the X-basis comparable to that of a standard laser. We have shown how the early, late and the superposition states can be experimentally generated, reaching an extinction ratio between the Z-basis states of about $25\,$dB. 
Furthermore, we measured the $g^2(0)$ auto-correlation function of the SLED, and derived from it the effective number of modes in each pulse, which is necessary to define the photon number emission probability of the source. 
In doing so, we evaluated the expected performance of a decoy-state BB84 protocol adopting this source. 
Simulations show that the multimode thermal statistics introduce only a small penalty compared to an ideal phase-randomized Poissonian source.

Future directions include the demonstration of the source in a complete QKD link, in order to measure the real-time QBER, SKR, and long-term stability in practical use-case conditions. 
Moreover, performance and applicability of this source could be enhanced by means of a complete finite-key security analysis for decoy-state QKD, explicitly implementing analytical bounds on the multimode thermal distribution. Such analysis could include a conservative lower bound on the effective number of modes that can be directly related to standard, convenient experimental tests. 
Additionally, one can investigate the possibility of gain-switching the SLED, as opposed to pulse-carving (some preliminary investigations of this are included in the SM, achieving $\sim 300$~ps pulse widths).
Finally, we believe that characterization of the photon-number distribution using photon-number-resolving detectors~\cite{cheng2022, stasi2024} or higher-order autocorrelation measurements would also provide additional value, both to confirm the statistical model and to provide an alternative method to compute the number of modes in each pulse. 
The comprehensive characterization and on-field validation of the technique will ultimately determine the scalability of SLED-based phase-correlation-free QKD systems operating at gigahertz rates.

\section*{Acknowledgements}

The authors thank M. Wu, S. Juárez and B. Taylor for insightful discussions. This work was supported by the European Union’s Horizon Europe Framework Programme under the Marie Sklodowska-Curie Grant No. 101072637 (Quantum-Safe Internet project); the European Space Agency (ESA) through the Phi-Lab Switzerland framework (MQ-SoC project); the Swiss State Secretariat for Research and Innovation (SERI) (Contract UeM019-3); the Galician Regional Government (consolidation of research units: atlanTTic); the Spanish Ministry Science, Innovation and Universities (MICIU); the Fondo Europeo de Desarrollo Regional (FEDER) through the grant No. PID2024-162270OB-I00; the “Hub Nacional de Excelencia en Comunicaciones Cuanticas” funded by the Spanish Ministry for Digital Transformation and the Public Service and the European Union NextGenerationEU; the project “Quantum Secure Networks Partnership” (QSNP, grant agreement No 101114043); the project IberianQCI (grant 101249593); as well as the Programa de Cooperación Interreg VI-A España–Portugal (POCTEP) 2021–2027 through the project QUANTUM\_IBER\_IA. 

\bibliography{paperpile}

@ARTICLE{mayers2001,
  title     = "{Unconditional security in quantum cryptography}",
  author    = "Mayers, Dominic",
  journal   = "J. ACM",
  publisher = "Association for Computing Machinery (ACM)",
  volume    =  48,
  number    =  3,
  pages     = "351-406",
  year      =  2001,
  url       = "http://dx.doi.org/10.1145/382780.382781"
}

@ARTICLE{pereira2020,
  title     = "{Quantum key distribution with correlated sources}",
  author    = "Pereira, Margarida and Kato, Go and Mizutani, Akihiro and Curty,
               Marcos and Tamaki, Kiyoshi",
  journal   = "Sci. Adv.",
  publisher = "American Association for the Advancement of Science (AAAS)",
  volume    =  6,
  number    =  37,
  pages     = "eaaz4487",
  year      =  2020,
  url       = "http://dx.doi.org/10.1126/sciadv.aaz4487"
}

@ARTICLE{aquina2025,
  title   = "{A critical analysis of deployed use cases for quantum key
             distribution and comparison with post-quantum cryptography}",
  author  = "Aquina, Nick and Cimoli, Bruno and Das, Soumya and H{\"o}velmanns,
             Kathrin and Weber, Fiona Johanna and Okonkwo, Chigo and Rommel,
             Simon and {\v S}kori{\'c}, Boris and Tafur Monroy, Idelfonso and Verschoor,
             Sebastian",
  journal = "EPJ Quantum Technol.",
  volume  =  12,
  number  =  1,
  pages   =  51,
  year    =  2025,
  url     = "http://dx.doi.org/10.1140/epjqt/s40507-025-00350-5"
}

@ARTICLE{yin2020b,
  title     = "{Entanglement-based secure quantum cryptography over 1,120
               kilometres}",
  author    = "Yin, Juan and Li, Yu-Huai and Liao, Sheng-Kai and Yang, Meng and
               Cao, Yuan and Zhang, Liang and Ren, Ji-Gang and Cai, Wen-Qi and
               Liu, Wei-Yue and Li, Shuang-Lin and Shu, Rong and Huang, Yong-Mei
               and Deng, Lei and Li, Li and Zhang, Qiang and Liu, Nai-Le and
               Chen, Yu-Ao and Lu, Chao-Yang and Wang, Xiang-Bin and Xu, Feihu
               and Wang, Jian-Yu and Peng, Cheng-Zhi and Ekert, Artur K and Pan,
               Jian-Wei",
  journal   = "Nature",
  publisher = "Springer Science and Business Media LLC",
  volume    =  582,
  number    =  7813,
  pages     = "501-505",
  year      =  2020,
  url       = "http://dx.doi.org/10.1038/s41586-020-2401-y"
}

@ARTICLE{li2023,
  title     = "{High-rate quantum key distribution exceeding 110 Mb s--1}",
  author    = "Li, Wei and Zhang, Likang and Tan, Hao and Lu, Yichen and Liao,
               Sheng-Kai and Huang, Jia and Li, Hao and Wang, Zhen and Mao,
               Hao-Kun and Yan, Bingze and Li, Qiong and Liu, Yang and Zhang,
               Qiang and Peng, Cheng-Zhi and You, Lixing and Xu, Feihu and Pan,
               Jian-Wei",
  journal   = "Nat. Photonics",
  publisher = "Nature Publishing Group",
  volume    =  17,
  number    =  5,
  pages     = "416-421",
  year      =  2023,
  url       = "https://www.nature.com/articles/s41566-023-01166-4"
}

@ARTICLE{grunenfelder2023,
  title   = "{Fast single-photon detectors and real-time key distillation enable
             high secret-key-rate quantum key distribution systems}",
  author  = "Gr{\"u}nenfelder, Fadri and Boaron, Alberto and Resta, Giovanni V and
             Perrenoud, Matthieu and Rusca, Davide and Barreiro, Claudio and
             Houlmann, Rapha{\"e}l and Sax, Rebecka and Stasi, Lorenzo and
             El-Khoury, Sylvain and H{\"a}nggi, Esther and Bosshard, Nico and
             Bussi{\`e}res, F{\'e}lix and Zbinden, Hugo",
  journal = "Nat. Photonics",
  volume  =  17,
  number  =  5,
  pages   = "422-426",
  year    =  2023,
  url     = "http://dx.doi.org/10.1038/s41566-023-01168-2"
}

@ARTICLE{xu2020,
  title     = "{Secure quantum key distribution with realistic devices}",
  author    = "Xu, Feihu and Ma, Xiongfeng and Zhang, Qiang and Lo, Hoi-Kwong
               and Pan, Jian-Wei",
  journal   = "Rev. Mod. Phys.",
  publisher = "American Physical Society (APS)",
  volume    =  92,
  number    =  2,
  year      =  2020,
  url       = "http://dx.doi.org/10.1103/revmodphys.92.025002"
}

@ARTICLE{brazaola-vicario2024,
  title     = "{Quantum key distribution: a survey on current vulnerability
               trends and potential implementation risks}",
  author    = "Brazaola-Vicario, Aitor and Ruiz, Alejandra and Lage, Oscar and
               Jacob, Eduardo and Astorga, Jasone",
  journal   = "Opt. Contin.",
  publisher = "Optica Publishing Group",
  volume    =  3,
  number    =  8,
  pages     =  1438,
  year      =  2024,
  url       = "http://dx.doi.org/10.1364/optcon.530352"
}

@INPROCEEDINGS{bennett1984,
  title     = "{Quantum cryptography: Public key distribution and coin tossing}",
  author    = "Bennett, Charles H and Brassard, Gilles",
  booktitle = "{Proceedings of the IEEE International Conference on Computers,
               Systems and Signal Processing}",
  publisher = "IEEE",
  pages     = "175-179",
  year      =  1984
}

@ARTICLE{cheng2022,
  title     = "{A 100-pixel photon-number-resolving detector unveiling photon
               statistics}",
  author    = "Cheng, Risheng and Zhou, Yiyu and Wang, Sihao and Shen, Mohan and
               Taher, Towsif and Tang, Hong X",
  journal   = "Nat. Photonics",
  publisher = "Nature Publishing Group",
  volume    =  17,
  number    =  1,
  pages     = "112-119",
  year      =  2022,
  url       = "https://www.nature.com/articles/s41566-022-01119-3"
}

@ARTICLE{hwang2003,
  title     = "{Quantum key distribution with high loss: toward global secure
               communication}",
  author    = "Hwang, Won-Young",
  journal   = "Phys. Rev. Lett.",
  publisher = "American Physical Society (APS)",
  volume    =  91,
  number    =  5,
  pages     =  057901,
  year      =  2003,
  url       = "http://dx.doi.org/10.1103/PhysRevLett.91.057901"
}

@ARTICLE{tupkary2025,
  title   = "{QKD security proofs for decoy-state BB84: protocol variations,
             proof techniques, gaps and limitations}",
  author  = "Tupkary, Devashish and Tan, Ernest Y-Z and Nahar, Shlok and Kamin,
             Lars and L{\"u}tkenhaus, Norbert",
  journal = "arXiv [quant-ph]",
  year    =  2025,
  url     = "http://arxiv.org/abs/2502.10340"
}

@ARTICLE{stasi2024,
  title     = "{Enhanced detection rate and high photon-number efficiencies with
               a scalable parallel {SNSPD}}",
  author    = "Stasi, Lorenzo and Taher, Towsif and Resta, Giovanni V and
               Zbinden, Hugo and Thew, Rob and Bussi{\`e}res, F{\'e}lix",
  journal   = "ACS Photonics",
  publisher = "American Chemical Society (ACS)",
  volume    =  12,
  pages     = "320--329",
  year      =  2024,
  url       = "https://scholar.google.com/citations?view_op=view_citation&hl=en&citation_for_view=h5oKA0EAAAAJ:zYLM7Y9cAGgC"
}

@ARTICLE{navarrete2026,
  title   = "{Numerical security analysis for practical quantum key
             distribution}",
  author  = "Navarrete, {\'A}lvaro and Curr{\'a}s-Lorenzo, Guillermo and Pereira,
             Margarida and Curty, Marcos",
  journal = "arXiv:2605.12984 [quant-ph]",
  year    =  2026,
  url     = "http://dx.doi.org/10.48550/arXiv.2605.12984"
}

@ARTICLE{tittel2025,
  title     = "{Quantum networks using rare-earth ions}",
  author    = "Tittel, Wolfgang and Afzelius, Mikael and Kinos, A and Rippe, L
               and Walther, A",
  journal   = "Quantum Sci Technol",
  publisher = "IOP Publishing",
  volume    =  10,
  number    =  3,
  pages     =  033002,
  year      =  2025,
  url       = "https://scholar.google.com/citations?view_op=view_citation&hl=en&citation_for_view=gMESdFEAAAAJ:wKETBy42zhYC"
}

@INPROCEEDINGS{chen2017c,
  title     = "{Multi-channel high speed quantum random number generating with
               DWDM and superluminescent {LED}}",
  author    = "Chen, Ziyang and Wang, Gan and Li, Zhengyu and Peng, Xiang and
               Guo, Hong",
  booktitle = "{2017 IEEE 85th Vehicular Technology Conference (VTC Spring)}",
  publisher = "IEEE",
  year      =  2017,
  url       = "http://dx.doi.org/10.1109/vtcspring.2017.8108695"
}

@ARTICLE{yang2021,
  title     = "{Randomness quantification for quantum random number generation
               based on detection of amplified spontaneous emission noise}",
  author    = "Yang, Jie and Fan, Fan and Liu, Jinlu and Su, Qi and Li, Yang and
               Huang, Wei and Xu, Bingjie",
  journal   = "Quantum Sci. Technol.",
  publisher = "IOP Publishing",
  volume    =  6,
  number    =  1,
  pages     =  015002,
  year      =  2021,
  url       = "http://dx.doi.org/10.1088/2058-9565/abbd80"
}

@ARTICLE{pietralunga2003,
  title     = "{Photon statistics of amplified spontaneous emission in a dense
               wavelength-division multiplexing regime}",
  author    = "Pietralunga, Silvia M and Martelli, Paolo and Martinelli, Mario",
  journal   = "Opt. Lett.",
  publisher = "Optica Publishing Group",
  volume    =  28,
  number    =  3,
  pages     = "152-154",
  year      =  2003,
  url       = "http://dx.doi.org/10.1364/ol.28.000152"
}

@ARTICLE{wong1998,
  title     = "{Photon statistics of amplified spontaneous emission noise in a
               10-Gbit/s optically preamplified direct-detection receiver}",
  author    = "Wong, W S and Haus, H A and Jiang, L A and Hansen, P B and
               Margalit, M",
  journal   = "Opt. Lett.",
  publisher = "Optica Publishing Group",
  volume    =  23,
  number    =  23,
  pages     = "1832-1834",
  year      =  1998,
  url       = "http://dx.doi.org/10.1364/ol.23.001832"
}

@ARTICLE{li2021,
  title     = "{Experimental study on the security of superluminescent LED-based
               quantum random generator}",
  author    = "Li, Yuanhao and Fei, Yangyang and Wang, Weilong and Meng,
               Xiangdong and Wang, Hong and Duan, Qianheng and Ma, Zhi",
  journal   = "Opt. Eng.",
  publisher = "SPIE-Intl Soc Optical Eng",
  volume    =  60,
  number    =  11,
  pages     =  116106,
  year      =  2021,
  url       = "http://dx.doi.org/10.1117/1.oe.60.11.116106"
}

@ARTICLE{lo2026,
  title   = "{Phase-randomized laser pulse generation at 10 GHz for quantum
             photonic applications}",
  author  = "Lo, Yuen San and Brzosko, Adam H and Smith, Peter R and Woodward,
             Robert I and Marangon, Davide G and Dynes, James F and Ju{\'a}rez,
             Sergio and Para{\"i}so, Taofiq K and Stevenson, R Mark and Shields,
             Andrew J",
  journal = "arXiv:2601.04031 [quant-ph]",
  year    =  2026,
  url     = "http://dx.doi.org/10.48550/arXiv.2601.04031"
}

@ARTICLE{lo2014,
  title     = "{Secure quantum key distribution}",
  author    = "Lo, Hoi-Kwong and Curty, Marcos and Tamaki, Kiyoshi",
  journal   = "Nat. Photonics",
  publisher = "Springer Science and Business Media LLC",
  volume    =  8,
  number    =  8,
  pages     = "595-604",
  year      =  2014,
  url       = "http://dx.doi.org/10.1038/nphoton.2014.149"
}

@ARTICLE{pirandola2020b,
  title     = "{Advances in quantum cryptography}",
  author    = "Pirandola, S and Andersen, U L and Banchi, L and Berta, M and
               Bunandar, D and Colbeck, R and Englund, D and Gehring, T and
               Lupo, C and Ottaviani, C and Pereira, J L and Razavi, M and
               Shamsul Shaari, J and Tomamichel, M and Usenko, V C and Vallone,
               G and Villoresi, P and Wallden, P",
  journal   = "Adv. Opt. Photonics",
  publisher = "Optica Publishing Group",
  volume    =  12,
  number    =  4,
  pages     =  1012,
  year      =  2020,
  url       = "http://dx.doi.org/10.1364/AOP.361502"
}

@ARTICLE{ekert1991b,
  title     = "{Quantum cryptography based on Bell's theorem}",
  author    = "Ekert, A K",
  journal   = "Phys. Rev. Lett.",
  publisher = "American Physical Society (APS)",
  volume    =  67,
  number    =  6,
  pages     = "661-663",
  year      =  1991,
  url       = "http://dx.doi.org/10.1103/PhysRevLett.67.661"
}

@ARTICLE{lucamarini2018b,
  title     = "{Overcoming the rate-distance limit of quantum key distribution
               without quantum repeaters}",
  author    = "Lucamarini, M and Yuan, Z L and Dynes, J F and Shields, A J",
  journal   = "Nature",
  publisher = "Nature Publishing Group",
  volume    =  557,
  number    =  7705,
  pages     = "400-403",
  year      =  2018,
  url       = "http://dx.doi.org/10.1038/s41586-018-0066-6"
}

@ARTICLE{foletto2022,
  title     = "{Security bounds for decoy-state quantum key distribution with
               arbitrary photon-number statistics}",
  author    = "Foletto, Giulio and Picciariello, Francesco and Agnesi,
               Costantino and Villoresi, Paolo and Vallone, Giuseppe",
  journal   = "Phys. Rev. A",
  publisher = "American Physical Society (APS)",
  volume    =  105,
  number    =  1,
  pages     =  012603,
  year      =  2022,
  url       = "https://journals.aps.org/pra/abstract/10.1103/PhysRevA.105.012603"
}

@ARTICLE{shor2000,
  title     = "{Simple proof of security of the BB84 quantum key distribution
               protocol}",
  author    = "Shor, P W and Preskill, J",
  journal   = "Phys. Rev. Lett.",
  publisher = "American Physical Society (APS)",
  volume    =  85,
  number    =  2,
  pages     = "441-444",
  year      =  2000,
  url       = "http://dx.doi.org/10.1103/PhysRevLett.85.441"
}

@BOOK{loudon2000,
  title     = "{The quantum theory of light}",
  author    = "Loudon, Rodney",
  publisher = "Oxford University Press",
  edition   =  3,
  year      =  2000,
  url       = "https://www.amazon.com/Quantum-Theory-Oxford-Science-Publications/dp/0198501765"
}

@ARTICLE{mandel1996,
  title     = "{Optical coherence and quantum optics}",
  author    = "Mandel, Leonard and Wolf, Emil and Shapiro, Jeffrey H",
  journal   = "Am. J. Phys.",
  publisher = "American Institute of Physics",
  volume    =  64,
  pages     = "1438--1439",
  year      =  1996,
  url       = "https://pubs.aip.org/physicstoday/article-pdf/49/5/68/8309649/68_2_online.pdf"
}

@ARTICLE{dynes2018,
  title     = "{Testing the photon-number statistics of a quantum key
               distribution light source}",
  author    = "Dynes, J F and Lucamarini, M and Patel, K A and Sharpe, A W and
               Ward, M B and Yuan, Z L and Shields, A J",
  journal   = "Opt. Express",
  publisher = "Optica Publishing Group",
  volume    =  26,
  number    =  18,
  pages     = "22733-22749",
  year      =  2018,
  url       = "http://dx.doi.org/10.1364/OE.26.022733"
}

@ARTICLE{lo2005b,
  title     = "{Decoy state quantum key distribution}",
  author    = "Lo, Hoi-Kwong and Ma, Xiongfeng and Chen, Kai",
  journal   = "Phys. Rev. Lett.",
  publisher = "American Physical Society (APS)",
  volume    =  94,
  number    =  23,
  pages     =  230504,
  year      =  2005,
  url       = "http://dx.doi.org/10.1103/PhysRevLett.94.230504"
}

@ARTICLE{ma2005b,
  title     = "{Practical decoy state for quantum key distribution}",
  author    = "Ma, Xiongfeng and Qi, Bing and Zhao, Yi and Lo, Hoi-Kwong",
  journal   = "Phys. Rev. A",
  publisher = "American Physical Society (APS)",
  volume    =  72,
  number    =  1,
  pages     =  012326,
  year      =  2005,
  url       = "http://dx.doi.org/10.1103/PhysRevA.72.012326"
}

@ARTICLE{tamaki2014b,
  title     = "{Loss-tolerant quantum cryptography with imperfect sources}",
  author    = "Tamaki, Kiyoshi and Curty, Marcos and Kato, Go and Lo, Hoi-Kwong
               and Azuma, Koji",
  journal   = "Phys. Rev. A",
  publisher = "American Physical Society (APS)",
  volume    =  90,
  number    =  5,
  pages     =  052314,
  year      =  2014,
  url       = "http://dx.doi.org/10.1103/PhysRevA.90.052314"
}

@TECHREPORT{bsi2023,
  title       = "{A Study on Implementation Attacks against QKD Systems}",
  author      = "{BSI}",
  institution = "Federal Office for Information Security",
  year        =  2023,
  url         = "https://www.bsi.bund.de/EN/Service-Navi/Publikationen/Studien/QKD-Systems/Implementation_Attacks_QKD_Systems_node.html"
}

@ARTICLE{briegel1998b,
  title     = "{Quantum repeaters: The role of imperfect local operations in
               quantum communication}",
  author    = "Briegel, H-J and D{\"u}r, W and Cirac, J I and Zoller, P",
  journal   = "Phys. Rev. Lett.",
  publisher = "American Physical Society (APS)",
  volume    =  81,
  number    =  26,
  pages     = "5932-5935",
  year      =  1998,
  url       = "http://dx.doi.org/10.1103/physrevlett.81.5932"
}

@ARTICLE{azuma2023b,
  title     = "{Quantum repeaters: From quantum networks to the quantum
               internet}",
  author    = "Azuma, Koji and Economou, Sophia E and Elkouss, David and
               Hilaire, Paul and Jiang, Liang and Lo, Hoi-Kwong and Tzitrin,
               Ilan",
  journal   = "Rev. Mod. Phys.",
  publisher = "American Physical Society (APS)",
  volume    =  95,
  number    =  4,
  pages     =  045006,
  year      =  2023,
  url       = "http://dx.doi.org/10.1103/RevModPhys.95.045006"
}

@ARTICLE{makarov2024,
  title     = "{Preparing a commercial quantum key distribution system for
               certification against implementation loopholes}",
  author    = "Makarov, Vadim and Abrikosov, Alexey and Chaiwongkhot, Poompong
               and Fedorov, Aleksey K and Huang, Anqi and Kiktenko, Evgeny and
               Petrov, Mikhail and Ponosova, Anastasiya and Ruzhitskaya, Daria
               and Tayduganov, Andrey and Trefilov, Daniil and Zaitsev,
               Konstantin",
  journal   = "Phys. Rev. Appl.",
  publisher = "American Physical Society (APS)",
  volume    =  22,
  number    =  4,
  pages     =  044076,
  year      =  2024,
  url       = "http://dx.doi.org/10.1103/PhysRevApplied.22.044076"
}

@ARTICLE{sixto2025,
  title     = "{Quantum key distribution with imperfectly isolated devices}",
  author    = "Sixto, Xoel and Navarrete, {\'A}lvaro and Pereira, Margarida and
               Curr{\'a}s-Lorenzo, Guillermo and Tamaki, Kiyoshi and Curty, Marcos",
  journal   = "Quantum Sci. Technol.",
  publisher = "IOP Publishing",
  volume    =  10,
  number    =  3,
  pages     =  035034,
  year      =  2025,
  url       = "http://dx.doi.org/10.1088/2058-9565/addb6e"
}

@ARTICLE{zapatero2025b,
  title     = "{Implementation security in quantum key distribution}",
  author    = "Zapatero, V{\'i}ctor and Navarrete, {\'A}lvaro and Curty, Marcos",
  journal   = "Adv. Quantum Technol.",
  publisher = "Wiley",
  volume    =  8,
  number    =  2,
  pages     =  2300380,
  year      =  2025,
  url       = "https://d-nb.info/1315355590/34"
}

@ARTICLE{agulleiro2025,
  title   = "{Modeling and characterization of arbitrary order pulse
             correlations for quantum key distribution}",
  author  = "Agulleiro, Ainhoa and Gr{\"u}nenfelder, Fadri and Pereira, Margarida
             and Curr{\'a}s-Lorenzo, Guillermo and Zbinden, Hugo and Curty, Marcos
             and Rusca, Davide",
  journal = "arXiv [quant-ph] 2506.18684",
  year    =  2025,
  url     = "http://dx.doi.org/10.48550/arXiv.2506.18684"
}

@ARTICLE{curras-lorenzo2026,
  title   = "{Rigorous phase-error-estimation security framework for QKD with
             correlated sources}",
  author  = "Curr{\'a}s-Lorenzo, Guillermo and Pereira, Margarida and Tamaki,
             Kiyoshi and Curty, Marcos",
  journal = "arXiv [quant-ph] 2601.08417",
  year    =  2026,
  url     = "http://dx.doi.org/10.48550/arXiv.2601.08417"
}

@ARTICLE{pereira2024,
  title     = "{Quantum key distribution with unbounded pulse correlations}",
  author    = "Pereira, Margarida and Curr{\'a}s-Lorenzo, Guillermo and Mizutani,
               Akihiro and Rusca, Davide and Curty, Marcos and Tamaki, Kiyoshi",
  journal   = "Quantum Sci. Technol.",
  publisher = "IOP Publishing",
  volume    =  10,
  pages     =  015001,
  year      =  2024,
  url       = "https://iopscience.iop.org/article/10.1088/2058-9565/ad8181"
}

@ARTICLE{marcomini2025c,
  title     = "{Characterising higher-order phase correlations in gain-switched
               laser sources with application to quantum key distribution}",
  author    = "Marcomini, Alessandro and Curr{\'a}s-Lorenzo, Guillermo and Rusca,
               Davide and Valle, Angel and Tamaki, Kiyoshi and Curty, Marcos",
  journal   = "EPJ Quantum Technol.",
  publisher = "Springer Science and Business Media LLC",
  volume    =  12,
  number    =  1,
  pages     =  38,
  year      =  2025,
  url       = "https://link.springer.com/article/10.1140/epjqt/s40507-025-00340-7"
}

@ARTICLE{trefilov2025,
  title     = "{Intensity correlations in decoy-state BB84 quantum key
               distribution systems}",
  author    = "Trefilov, Daniil and Sixto, Xoel and Zapatero, V{\'i}ctor and Huang,
               Anqi and Curty, Marcos and Makarov, Vadim",
  journal   = "Opt. Quantum",
  publisher = "Optica Publishing Group",
  volume    =  3,
  number    =  5,
  pages     =  417,
  year      =  2025,
  url       = "http://dx.doi.org/10.1364/OPTICAQ.549690"
}

@ARTICLE{curras-lorenzo2024,
  title     = "{Security of quantum key distribution with imperfect phase
               randomisation}",
  author    = "Curr{\'a}s-Lorenzo, Guillermo and Nahar, Shlok and L{\"u}tkenhaus,
               Norbert and Tamaki, Kiyoshi and Curty, Marcos",
  journal   = "Quantum Sci. Technol.",
  publisher = "IOP Publishing",
  volume    =  9,
  number    =  1,
  pages     =  015025,
  year      =  2024,
  url       = "http://dx.doi.org/10.1088/2058-9565/ad141c"
}

@ARTICLE{boaron2018d,
  title     = "{Simple 2.5 GHz time-bin quantum key distribution}",
  author    = "Boaron, Alberto and Korzh, Boris and Houlmann, Raphael and Boso,
               Gianluca and Rusca, Davide and Gray, Stuart and Li, Ming-Jun and
               Nolan, Daniel and Martin, Anthony and Zbinden, Hugo",
  journal   = "Appl. Phys. Lett.",
  publisher = "AIP Publishing",
  volume    =  112,
  number    =  17,
  pages     =  171108,
  year      =  2018,
  url       = "http://dx.doi.org/10.1063/1.5027030"
}

@ARTICLE{zapatero2021,
  title     = "{Security of quantum key distribution with intensity
               correlations}",
  author    = "Zapatero, V{\'i}ctor and Navarrete, {\'A}lvaro and Tamaki, Kiyoshi and
               Curty, Marcos",
  journal   = "Quantum",
  publisher = "Verein zur Forderung des Open Access Publizierens in den
               Quantenwissenschaften",
  volume    =  5,
  number    =  602,
  pages     =  602,
  year      =  2021,
  url       = "http://dx.doi.org/10.22331/q-2021-12-07-602"
}

@ARTICLE{chen2021b,
  title     = "{An integrated space-to-ground quantum communication network over
               4,600 kilometres}",
  author    = "Chen, Yu-Ao and Zhang, Qiang and Chen, Teng-Yun and Cai, Wen-Qi
               and Liao, Sheng-Kai and Zhang, Jun and Chen, Kai and Yin, Juan
               and Ren, Ji-Gang and Chen, Zhu and Han, Sheng-Long and Yu, Qing
               and Liang, Ken and Zhou, Fei and Yuan, Xiao and Zhao, Mei-Sheng
               and Wang, Tian-Yin and Jiang, Xiao and Zhang, Liang and Liu,
               Wei-Yue and Li, Yang and Shen, Qi and Cao, Yuan and Lu, Chao-Yang
               and Shu, Rong and Wang, Jian-Yu and Li, Li and Liu, Nai-Le and
               Xu, Feihu and Wang, Xiang-Bin and Peng, Cheng-Zhi and Pan,
               Jian-Wei",
  journal   = "Nature",
  publisher = "Springer Science and Business Media LLC",
  volume    =  589,
  number    =  7841,
  pages     = "214-219",
  year      =  2021,
  url       = "http://dx.doi.org/10.1038/s41586-020-03093-8"
}

@ARTICLE{liao2017b,
  title     = "{Satellite-to-ground quantum key distribution}",
  author    = "Liao, Sheng-Kai and Cai, Wen-Qi and Liu, Wei-Yue and Zhang, Liang
               and Li, Yang and Ren, Ji-Gang and Yin, Juan and Shen, Qi and Cao,
               Yuan and Li, Zheng-Ping and Li, Feng-Zhi and Chen, Xia-Wei and
               Sun, Li-Hua and Jia, Jian-Jun and Wu, Jin-Cai and Jiang, Xiao-Jun
               and Wang, Jian-Feng and Huang, Yong-Mei and Wang, Qiang and Zhou,
               Yi-Lin and Deng, Lei and Xi, Tao and Ma, Lu and Hu, Tai and
               Zhang, Qiang and Chen, Yu-Ao and Liu, Nai-Le and Wang, Xiang-Bin
               and Zhu, Zhen-Cai and Lu, Chao-Yang and Shu, Rong and Peng,
               Cheng-Zhi and Wang, Jian-Yu and Pan, Jian-Wei",
  journal   = "Nature",
  publisher = "Springer Science and Business Media LLC",
  volume    =  549,
  number    =  7670,
  pages     = "43-47",
  year      =  2017,
  url       = "http://dx.doi.org/10.1038/nature23655"
}

@ARTICLE{pittaluga2025,
  title     = "{Long-distance coherent quantum communications in deployed
               telecom networks}",
  author    = "Pittaluga, Mirko and Lo, Yuen San and Brzosko, Adam and Woodward,
               Robert I and Scalcon, Davide and Winnel, Matthew S and Roger,
               Thomas and Dynes, James F and Owen, Kim A and Ju{\'a}rez, Sergio and
               Rydlichowski, Piotr and Vicinanza, Domenico and Roberts, Guy and
               Shields, Andrew J",
  journal   = "Nature",
  publisher = "Nature Publishing Group",
  volume    =  640,
  number    =  8060,
  pages     = "911-917",
  year      =  2025,
  url       = "http://dx.doi.org/10.1038/s41586-025-08801-w"
}

@ARTICLE{zheng2026,
  title     = "{Large-scale quantum communication networks with integrated
               photonics}",
  author    = "Zheng, Yun and Wang, Hanyu and Jia, Xinyu and Huang, Jiahui and
               Yuan, Huihong and Zhai, Chonghao and Dai, Junhao and Shi, Jingbo
               and Zhang, Lei and Zhang, Xuguang and Zhuang, Minxue and Liu,
               Jinchang and Mao, Jun and Dai, Tianxiang and Fu, Zhaorong and
               Jiao, Yuqing and Shi, Yaocheng and Dai, Daoxin and Wang, Xingjun
               and Li, Yan and Gong, Qihuang and Yuan, Zhiliang and Chang, Lin
               and Wang, Jianwei",
  journal   = "Nature",
  publisher = "Springer Science and Business Media LLC",
  volume    =  651,
  number    =  8104,
  pages     = "68-75",
  year      =  2026,
  url       = "http://dx.doi.org/10.1038/s41586-026-10152-z"
}

@ARTICLE{wu2025,
  title     = "{Integration of quantum key distribution and high-throughput
               classical communications in field-deployed multi-core fibers}",
  author    = "Wu, Qi and Ribezzo, Domenico and Di Sciullo, Giammarco and
               Cocchi, Sebastiano and Ann Shaji, Divya and Alves Zischler, Lucas
               and Luis, Ruben and Serena, Paolo and Lasagni, Chiara and Bononi,
               Alberto and Hayashi, Tetsuya and Gagliano, Alessandro and
               Martelli, Paolo and Gatto, Alberto and Parolari, Paola and Boffi,
               Pierpaolo and Bacco, Davide and Zavatta, Alessandro and Zhu,
               Yixiao and Hu, Weisheng and Xu, Zhaopeng and Shtaif, Mark and
               Marotta, Andrea and Graziosi, Fabio and Mecozzi, Antonio and
               Antonelli, Cristian",
  journal   = "Light Sci. Appl.",
  publisher = "Nature Publishing Group",
  volume    =  14,
  number    =  1,
  pages     =  274,
  year      =  2025,
  url       = "http://dx.doi.org/10.1038/s41377-025-01982-z"
}

@ARTICLE{lim2014d,
  title     = "{Concise security bounds for practical decoy-state quantum key
               distribution}",
  author    = "Lim, Charles Ci Wen and Curty, Marcos and Walenta, Nino and Xu,
               Feihu and Zbinden, Hugo",
  journal   = "Phys. Rev. A",
  publisher = "American Physical Society (APS)",
  volume    =  89,
  number    =  2,
  pages     =  022307,
  year      =  2014,
  url       = "http://dx.doi.org/10.1103/PhysRevA.89.022307"
}

@ARTICLE{rusca2018b,
  title     = "{Finite-key analysis for the 1-decoy state QKD protocol}",
  author    = "Rusca, Davide and Boaron, Alberto and Gr{\"u}nenfelder, Fadri and
               Martin, Anthony and Zbinden, Hugo",
  journal   = "Appl. Phys. Lett.",
  publisher = "AIP Publishing",
  volume    =  112,
  number    =  17,
  pages     =  171104,
  year      =  2018,
  url       = "http://dx.doi.org/10.1063/1.5023340"
}

@ARTICLE{grunenfelder2020b,
  title     = "{Performance and security of 5 GHz repetition rate
               polarization-based quantum key distribution}",
  author    = "Gr{\"u}nenfelder, Fadri and Boaron, Alberto and Rusca, Davide and
               Martin, Anthony and Zbinden, Hugo",
  journal   = "Appl. Phys. Lett.",
  publisher = "AIP Publishing",
  volume    =  117,
  number    =  14,
  pages     =  144003,
  year      =  2020,
  url       = "http://dx.doi.org/10.1063/5.0021468"
}

@ARTICLE{kobayashi2014,
  title     = "{Evaluation of the phase randomness of a light source in
               quantum-key-distribution systems with an attenuated laser}",
  author    = "Kobayashi, Toshiya and Tomita, Akihisa and Okamoto, Atsushi",
  journal   = "Phys. Rev. A",
  publisher = "American Physical Society (APS)",
  volume    =  90,
  number    =  3,
  pages     =  032320,
  year      =  2014,
  url       = "http://dx.doi.org/10.1103/PhysRevA.90.032320"
}

@ARTICLE{wang2005b,
  title     = "{Beating the photon-number-splitting attack in practical quantum
               cryptography}",
  author    = "Wang, Xiang-Bin",
  journal   = "Phys. Rev. Lett.",
  publisher = "American Physical Society (APS)",
  volume    =  94,
  number    =  23,
  pages     =  230503,
  year      =  2005,
  url       = "http://dx.doi.org/10.1103/PhysRevLett.94.230503"
}

@ARTICLE{scarani2009b,
  title     = "{The security of practical quantum key distribution}",
  author    = "Scarani, Valerio and Bechmann-Pasquinucci, Helle and Cerf,
               Nicolas J and Du{\v s}ek, Miloslav and L{\"u}tkenhaus, Norbert and Peev,
               Momtchil",
  journal   = "Rev. Mod. Phys.",
  publisher = "American Physical Society (APS)",
  volume    =  81,
  number    =  3,
  pages     = "1301-1350",
  year      =  2009,
  url       = "http://dx.doi.org/10.1103/revmodphys.81.1301"
}

@ARTICLE{gisin2002,
  title     = "{Quantum cryptography}",
  author    = "Gisin, Nicolas and Ribordy, Gr{\'e}goire and Tittel, Wolfgang and
               Zbinden, Hugo",
  journal   = "Rev. Mod. Phys.",
  publisher = "American Physical Society (APS)",
  volume    =  74,
  number    =  1,
  pages     = "145-195",
  year      =  2002,
  url       = "http://dx.doi.org/10.1103/revmodphys.74.145"
}

@ARTICLE{brassard2000,
  title     = "{Limitations on practical quantum cryptography}",
  author    = "Brassard, G and Lutkenhaus, N and Mor, T and Sanders, B C",
  journal   = "Phys. Rev. Lett.",
  publisher = "American Physical Society (APS)",
  volume    =  85,
  number    =  6,
  pages     = "1330-1333",
  year      =  2000,
  url       = "http://dx.doi.org/10.1103/PhysRevLett.85.1330"
}

@ARTICLE{zapatero2023c,
  title     = "{A fully passive transmitter for decoy-state quantum key
               distribution}",
  author    = "Zapatero, V{\'i}ctor and Wang, Wenyuan and Curty, Marcos",
  journal   = "Quantum Sci. Technol.",
  publisher = "IOP Publishing",
  volume    =  8,
  number    =  2,
  pages     =  025014,
  year      =  2023,
  url       = "http://dx.doi.org/10.1088/2058-9565/acbc46"
}

@ARTICLE{wang2023b,
  title     = "{Fully passive quantum key distribution}",
  author    = "Wang, Wenyuan and Wang, Rong and Hu, Chengqiu and Zapatero,
               Victor and Qian, Li and Qi, Bing and Curty, Marcos and Lo,
               Hoi-Kwong",
  journal   = "Phys. Rev. Lett.",
  publisher = "American Physical Society (APS)",
  volume    =  130,
  number    =  22,
  pages     =  220801,
  year      =  2023,
  url       = "http://dx.doi.org/10.1103/PhysRevLett.130.220801"
}

\clearpage
\onecolumngrid
\clearpage

\includepdf[
    pages={1},
    pagecommand={\thispagestyle{empty}},
    fitpaper=true,
    turn=false
]{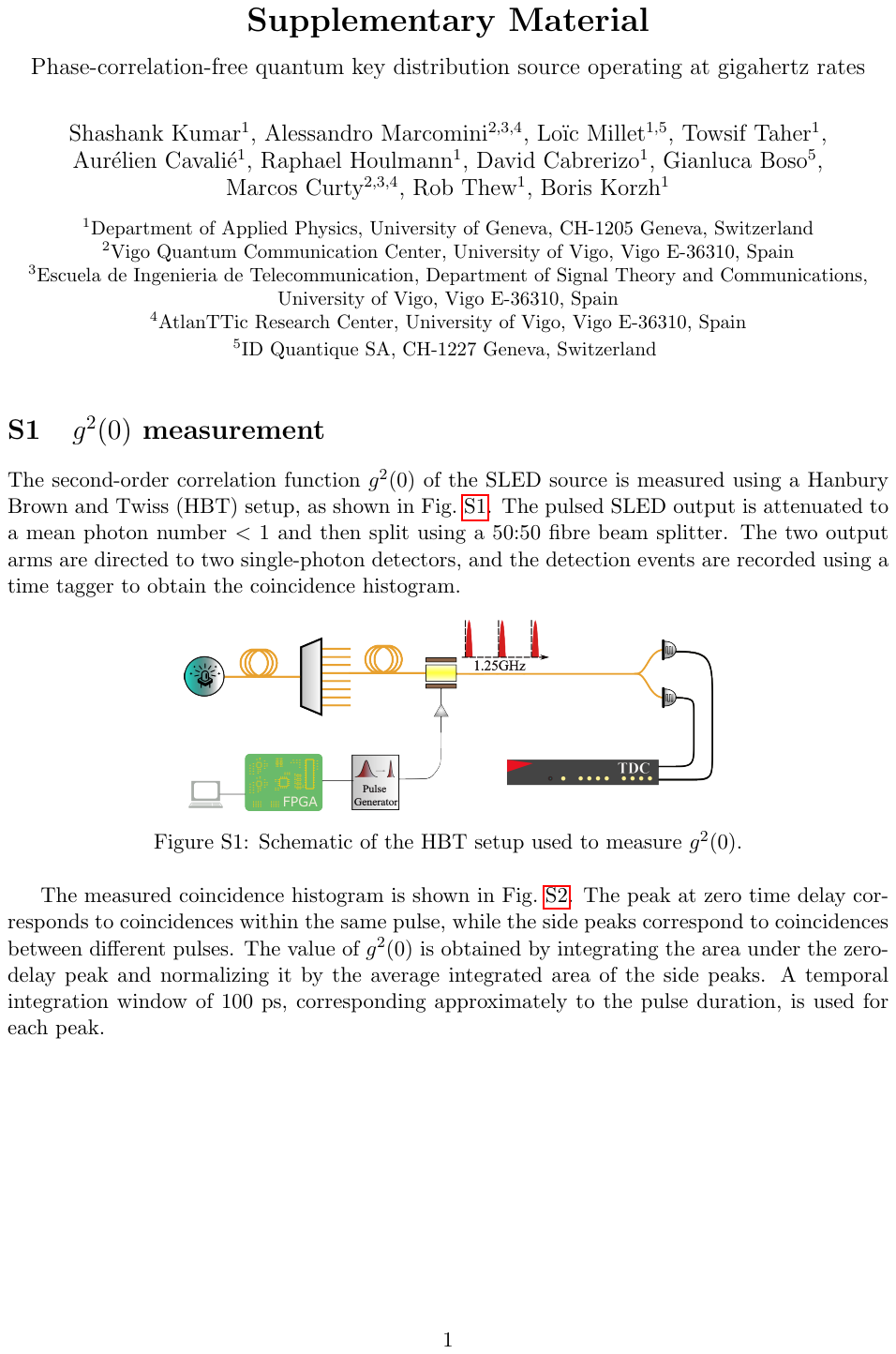}

\clearpage

\includepdf[
    pages={2},
    pagecommand={\thispagestyle{empty}},
    fitpaper=true,
    turn=false
]{Final/Supplymentary.pdf}

\clearpage
\includepdf[
    pages={3},
    pagecommand={\thispagestyle{empty}},
    fitpaper=true,
    turn=false
]{Final/Supplymentary.pdf}

\clearpage
\includepdf[
    pages={4},
    pagecommand={\thispagestyle{empty}},
    fitpaper=true,
    turn=false
]{Final/Supplymentary.pdf}

\end{document}